\documentclass[aps,twocolumn,nofootinbib]{revtex4}

\usepackage{amsmath,amssymb}
\usepackage{graphicx,color}
\newcommand{\amp}{&\!\!}
\newcommand{\mpl}{M_{\mbox{\tiny{Pl}}}}
\newcommand{\beq}{\begin{equation}}
\newcommand{\eeq}{\end{equation}}
\newcommand{\bea}{\begin{eqnarray}}
\newcommand{\eea}{\end{eqnarray}}
\newcommand{\newfi}{g}
\newcommand{\heafi}{\phi}
\newcommand{\heafiC}{\Phi}

\begin{document}

\title{{\bf Can Inflation be Connected to Low Energy Particle Physics?}}

\author{Mark~P.~Hertzberg\\
{\em Dept.~of Physics, Stanford University, Stanford, CA 94305, USA}}



\begin{abstract}
It is an interesting question whether low energy degrees of freedom may be responsible for early universe inflation. 
To examine this, here we present a simple version of Higgs-inflation with minimal coupling to gravity and a quadratic inflationary potential.
This quantitatively differs from the popular non-minimally coupled models, although it is qualitatively similar.
In all such models, new heavy fields must enter in order for the theory to be well behaved in the UV.
We show that in all cases the Higgs self coupling $\lambda$ must be quite small in order to integrate out the heavy fields and use the resulting low energy effective field theory of the Higgs to describe inflation.
For moderately sized $\lambda$, the UV completion is required and will, in general, determine the inflationary regime.
We discuss the important issue of the arbitrariness of the Lagrangians used in all these setups by presenting a new class of such models, including a supergravity version. This suggests that the inflationary potential is  disconnected from low energy physics.
\end{abstract}

\maketitle  

\let\thefootnote\relax\footnotetext{\!Electronic address: mphertz@stanford.edu}



\section{Introduction}
A large effort is underway to unravel the physical mechanism behind cosmological inflation \cite{Guth,Linde,AlbrechtSteinhardt82}, since it is the paradigm description of the very early universe. In particular, this requires the discovery of the degrees of freedom and the interaction Lagrangian that underpins inflation. In most models, the inflationary energy scale is many orders of magnitude higher than that probed in colliders, such as the LHC. For instance, models often involve energies around, or close to, the GUT scale $\sim 10^{15-16}$\,GeV, while colliders probe much lower energies $\sim 10^{3-4}$\,GeV. It is conceivable that the degrees of freedom relevant at these lower energies are also responsible for the physics of inflation at much higher energies, though by no means is it assured or likely.

In order to address this issue, one must measure a Lagrangian at low energies and extrapolate its predictions to very high scales.
The principles of effective field theory suggest that at high scales new operators will be important and alter the physics. In the case of inflation this leads to a great ambiguity in the form of the inflationary Lagrangian, leading to countless different models.
In recent times it has become popular to assume a particular form for the Lagrangian, namely a standard model (or supersymmetric) Higgs field with quartic potential ${\lambda\over 4}h^4$ and non-minimal coupling to gravity ${1\over2}\xi h^2\mathcal{R}$, 
which achieves inflation for large field values $h\gtrsim\mpl/\sqrt{\xi}$ and $\xi\sim 10^{4-5}$ \cite{Salopek,Bezrukov:2007ep}. As explained in Ref.~\cite{DeSimone:2008ei} this model requires the Higgs mass to be in the range $126\,\mbox{GeV}\lesssim m_H\lesssim185$\,GeV for the standard model, although the lower bound can be relaxed in supersymmetric models. The latest collider data puts the Higgs mass in the range 
$114\,\mbox{GeV}\lesssim m_H\lesssim 145$\,GeV, so this remains a possibility. Also the model predicts $n_s\approx 0.97$ (with corrections near any regime of vacuum instability) which is compatible with the latest WMAP7 data \cite{Komatsu:2010fb}. 
However, the applicability of the low energy Lagrangian during inflation has come under scrutiny \cite{Burgess:2009ea,Hertzberg:2010dc}, since new physics must enter at a UV scale $\Lambda\sim\mpl/\xi$. So one might wonder if such a Lagrangian is the only unique form, or if alternatives are possible which may alter the predictions, or worse, that the new physics entirely changes the inflationary regime.
Some related work in this general area of study, with wide-ranging conclusions, includes Refs.~\cite{Nakayama:2010sk, Mazumdar:2011ih,Barbon:2009ya,Bezrukov:2009yw,Bezrukov:2010jz,Atkins:2010yg}.

In this paper we address these issues. We construct a simpler version of Higgs-inflation in which all fields are minimally coupled to gravity. In order to track any possible breakdown of the low energy effective field theory, we explicitly introduce a new field that UV completes the theory in a fashion similar to Ref.~\cite{Giudice:2010ka}. Our central results are rather analogous to Ref.~\cite{Giudice:2010ka}, but will be much simpler and more transparent. Indeed although our results are in a specific setting, the central qualitative conclusions will be of general applicability to all Higgs-inflation models, both minimally or non-minimally coupled.
In Section \ref{Simple} we present this simple model. 
In Section \ref{Effective} we integrate out  heavy fields whenever we can in order to construct an effective field theory for the Higgs to describe inflation.
We then generalize this construction in Section \ref{Generalization}. 
We embed the model into the framework of supergravity in Section \ref{Supergravity}. Finally, we present our conclusions in Section \ref{Conclusions}.

\section{Simple Model}\label{Simple}
Consider a pair of scalar fields: Higgs $h$ and a new heavy scalar $\heafi$. We could include the three Goldstone bosons of the standard model Higgs, which play an important role in the computation of scattering cross sections, as explained in \cite{Hertzberg:2010dc}, but will not alter our basic results here. The role of the second field $\heafi$ is to allow us to track the UV behavior of the theory more clearly.
For simplicity, let's minimally couple $h$ and $\heafi$ to gravity and consider the following Lagrangian that is manifestly dimension 4 in the scalar sectors (see Section \ref{Supergravity} for an embedding in supergravity)
\bea
S=\int d^4x\sqrt{-g}\!\!\!\!\!\!\!\amp\amp\Bigg{[}{1\over2}\mpl^2\mathcal{R}+{1\over2}(\partial h)^2-{\lambda\over4}h^4\nonumber\\
\amp+\amp{1\over 2}(\partial\heafi)^2-{\newfi\over4}\left(\sqrt{2\over3}{\mpl\over\xi}\heafi-h^2\right)^{\!2}\Bigg{]}\,\,\,\,\,
\label{action}\eea
with Planck mass $\mpl\equiv 1/\sqrt{8\pi G_N}$ and metric signature $(+1,-1,-1, -1)$.
If one were to compute scattering amplitudes or compute quantum corrections to the action, one would readily see that this action has a Planckian cutoff. Graviton exchange leads to an entire tower of derivative corrections that are suppressed by powers of the Planck scale $\sim 1/\mpl$. Since inflation occurs at energy densities that are several orders of magnitude below the Planck energy density, this leads to negligibly small corrections to our results, as is true in standard slow-roll inflationary models. In Section IIIA we will integrate out $\phi$ (allowed for $\lambda\ll g$), leading to a cut-off in the effective field theory of $\sim\mpl/\xi$, while the full theory here including $\phi$ is a possible UV completion up to the Planck scale (we will later see, however, that the details are rather arbitrary).
 
\begin{figure}[t]
\center{\includegraphics[width=8.2cm]{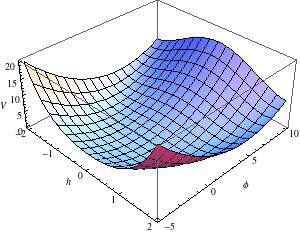}}
\caption{Potential $V$ as a function of $h$ and $\phi$.}
\label{PotentialPlot}\end{figure}

We have included the potential terms $h^4,\heafi^2,\heafi\, h^2$ and, without loss of generality, re-organized them into a convenient form in (\ref{action}).
A plot of the potential is given in Fig.~\ref{PotentialPlot}.
We could also include the dimension $\leq4$ terms $\heafi^3$, $\heafi^4$, $\heafi^2 h^2$, however, for simplicity, we assume them to be small (they are not generated at one-loop).
One should, in principle, alter the Higgs potential to ${\lambda\over 4}(h^2-v^2)^2$, where $v$ is the electroweak vev, but since $v\approx 246$\,GeV is so many order of magnitude lower than the scale of inflation, it can be ignored
(we shall not address the hierarchy problem for the Higgs mass in this paper, although see Section \ref{Supergravity} for supergravity).
The cross coupling is $\bar{m}\sim g\mpl/\xi$ and the $\heafi$ field has mass
\beq
m_\heafi=\sqrt{\newfi\over3}{\mpl\over\xi}
\label{sigmamass}\eeq
which is related to the scale of unitarity violation $\Lambda\sim\mpl/\xi$ \cite{Burgess:2009ea,Hertzberg:2010dc} in the $\heafi$-less theory by a factor of $\sim\sqrt{\newfi}$. One would normally expect the new physics to enter at a mass scale parametrically smaller than $\Lambda$, so we require $\newfi\lesssim1$, with $g\ll1$ preferable.
Also, the coefficient of $h^4$ is
\beq
\lambda_{tot}=\lambda+\newfi
\eeq
which means that neither $\lambda$ nor $\newfi$ can be large in order for the theory to be perturbative at high energies (in this paper, both parameters are taken to be positive). Also note that the $\sqrt{2/3}$ factor in the final term in (\ref{action}) is for convenience (so that $\xi$ here connects more directly with the $\xi$ that appears in the non-minimally coupled models).

\section{Effective Field Theory}\label{Effective}
At low energies we can integrate out the $\heafi$ field giving the low energy effective field theory for $h$. 
This comes from replacing $\heafi\to\sqrt{3\over2}{\xi\over\mpl} h^2$, plus kinetic corrections. At tree level we find
\bea
S_{\mbox{\tiny{EFT}}}=\int d^4x\sqrt{-g}\!\!\!\!\!\!\!\amp\amp\Bigg{[}{1\over2}\mpl^2\mathcal{R}+{1\over2}(\partial h)^2-{\lambda\over4}h^4\nonumber\\\amp+\amp{3\xi^2h^2\over\mpl^2}(\partial h)^2
+{18\xi^4\over\newfi\mpl^4}(\partial h)^4+\ldots\Bigg{]}\,\,\,\,\,\,
\label{EFTexp}\eea
Note that at low energies, the coefficient of $h^4$ is simply $\lambda$ and there is an infinite tower of kinetic corrections in the low energy effective field theory,
such as $\sim \newfi^{-1}(\xi/\mpl)^4(\partial h)^4$, in accord with the results of \cite{Hertzberg:2010dc}.

At large field values, the $\sim(\xi^2 h^2/\mpl^2)(\partial h)^2$ kinetic term is dominant. It is simple to check that in this regime, the only way for the higher order kinetic terms to be suppressed relative to the leading order terms is if $\lambda\ll\newfi$. Otherwise all the higher order operators matter and they will affect the inflationary phase. So lets consider the two limits in turn: (i) $\lambda\ll\newfi$ and (ii) $\newfi\ll\lambda$.

\subsection{Case (i) $\lambda\ll\newfi$}
Here we can safely ignore the higher order kinetic corrections in the low energy effective field theory, giving the following action for $h$
\bea
S_{\mbox{\tiny{EFT}}}=\int d^4x\sqrt{-g}\Bigg{[}\amp\amp\!\!\!\!\!\!\!\!{1\over2}\mpl^2\mathcal{R}+{1\over2}(\partial h)^2\nonumber\\
\amp-\amp{\lambda\over4}h^4+{3\xi^2h^2\over\mpl^2}(\partial h)^2\Bigg{]}
\label{smallL1}\eea
This action is identical to that used in the original Higgs-inflation model \cite{Salopek,Bezrukov:2007ep} (expressed in the Einstein frame), up to factors of $f(h)=(1+\xi h^2/\mpl^2)$. Such factors do not alter the regime of applicability (as this is controlled by the parametrically lower scale $\Lambda\sim\mpl/\xi$) but it does alter the shape of the inflationary potential and its cosmological predictions. The final term $\sim(\xi^2 h^2/\mpl^2)(\partial h^2)$ is present in all such models and exhibits the UV scale $\Lambda\sim\mpl/\xi$.

For $h\gg\mpl/\xi$, we can ignore the $\sim(\partial h)^2$ term relative to the $\sim(\xi^2 h^2/\mpl^2)(\partial h^2)$ term.
In this regime, it is best to switch back to the variable $\heafi$ as it carries canonical kinetic energy, giving
\beq
S_{\mbox{\tiny{EFT}}}\approx\int d^4x\sqrt{-g}\left[{1\over2}\mpl^2\mathcal{R}
-{\lambda\mpl^2\over6\,\xi^2}\heafi^2+{1\over2}(\partial \heafi)^2\right]
\label{smallL2}\eeq
This describes a model of quadratic chaotic inflation \cite{Linde:1983gd}.
The prediction for the spectral index is $n_s= 1-2/N_e\approx 0.96$, differing slightly from the non-minimally coupled model, which predicts $n_s\approx 0.97$ \cite{Bezrukov:2007ep}. 
The inflaton mass is given by
\beq
m_{\mbox{\tiny{inf}}}=\sqrt{\lambda\over3}{\mpl\over\xi}
\eeq
and the amplitude of density fluctuations in the chaotic inflation model is given by
\beq
\Delta_R^2={V(\phi)\over 24\,\pi^2\mpl^4\epsilon(\phi)}={m_{\mbox{\tiny{inf}}}^2\,N_e^2\over6\,\pi^2\mpl^6}
\eeq 
using $\phi=2\sqrt{N_e}\,\mpl$. Taking $N_e=55$ and demanding that we obtain the correct amplitude
of density fluctuations $\Delta_R^2\approx 2.4\times 10^{-9}$ \cite{Komatsu:2010fb}, we obtain the inflaton mass
$m_{\mbox{\tiny{inf}}}\approx 6.9\times 10^{-6}\mpl$. So this gives the constraint
\beq
{\lambda\over\xi^2}\approx 1.4\times 10^{-10}.
\eeq

Hence this provides a model of Higgs-inflation where $\lambda$ (as well as $\xi$) control the scale of inflation.
It is quite surprising that the inflationary model is that of a quadratic (or chaotic) model, rather than a quartic model, due to the $\sim h^2(\partial h)^2$ term that appears in the effective field theory.
However, it is important to emphasize that this requires $\lambda\ll\newfi$ in order to be valid. Furthermore, we know that  in order for high energy scattering of $h$'s to be perturbative, $\newfi$ cannot be large, and since we require $\newfi\lesssim1$, with $\newfi\ll1$ preferable, in order for $\heafi$ to enter at a scale parametrically lower than the cutoff of the effective field theory (as explained after eq.~(\ref{sigmamass})), this requires $\lambda$ to be quite small. 

\subsection{Case (ii) $\newfi\ll\lambda$}
In this case, we must keep track of the whole infinite tower of higher order operators in eq.~(\ref{EFTexp}). This puts us in a regime where the low energy effective field theory for $h$ breaks down. Instead we must know the details of the UV completion to study the inflationary physics; which means knowing the details of the $\heafi$ field interactions. In fact the physics of the heavy field $\heafi$ will be entirely responsible for the inflationary physics. In this regime, the dynamics of $h$ is largely irrelevant and we can essentially ignore its displacement from its $h=0$ (or electroweak) vev. So from our starting Lagrangian (with $h=0$) we immediately obtain
\beq
S=\int d^4x\sqrt{-g}\left[{1\over2}\mpl^2\mathcal{R}+{1\over 2}(\partial\heafi)^2
-{\newfi\mpl^2\over6\,\xi^2}\heafi^2\right]
\label{sigmaaction}\eeq
This describes a model of chaotic inflation with inflaton mass
\beq
m_{\mbox{\tiny{inf}}}=m_\heafi=\sqrt{\newfi\over3}{\mpl\over\xi}.
\eeq
So we have a new constraint involving $\newfi$
\beq
{\newfi\over\xi^2}\approx 1.4\times 10^{-10},
\label{constraintnewfi}\eeq
which is independent of the self coupling $\lambda$, i.e., inflation is essentially being driven by some hidden sector field $\heafi$ and not the Higgs.

\subsection{Generalization}\label{Generalization}
To emphasize the importance of the details of the new physics, we could start with a slightly more general action, namely
\bea
S\amp=\amp\int \! d^4x\sqrt{-g}\Bigg{[}{1\over2}\mpl^2\mathcal{R}+{1\over2}(\partial h)^2-{\lambda\over4}h^4\, G(h)
\nonumber\\
\amp\amp\,\,\,\,\,\,\,+{1\over 2}(\partial\heafi)^2
-{\newfi\over4}\left(\sqrt{2\over3}{\mpl\over\xi}\heafi-h^2\right)^{\!2}\!F\!\left(\heafi\right)\Bigg{]}\,\,\,\,
\label{actionG}\eea
where we have multiplied the Higgs potential and the last term by the dimensionless functions $G(h)$ and $F(\heafi)$, respectively.
These are assumed to satisfy $G(0)=F(0)=1$ and be slowly varying over field ranges of order $\Delta h\sim\mpl/\sqrt{\xi}$ for $G(h)$ and the Planck mass $\Delta\heafi\sim\mpl$ for $F(\heafi)$, or some other high scale. The previous model is recovered for $G(h)\equiv F(\heafi)\equiv1$. Note that this Lagrangian still has a very high scale cutoff. One is {\em allowed} to introduce other operators with sufficiently large coefficients to be relevant during the inflationary regime, however they are not {\em required}, since the quantum mechanically generated corrections are small. We shall therefore ignore such possibilities here.

For small $\lambda$ we can again integrate out the $\heafi$ field and obtain a modification of the chaotic inflation model of case (i) given in eq.~(\ref{smallL1}), namely
\bea
S_{\mbox{\tiny{EFT}}}=\int d^4x\sqrt{-g}\Bigg{[}\amp\amp\!\!\!\!\!\!\!\!{1\over2}\mpl^2\mathcal{R}+{1\over2}(\partial h)^2\nonumber\\
\amp-\amp{\lambda\over4}h^4 G(h)+{3\xi^2h^2\over\mpl^2}(\partial h)^2\Bigg{]}\,\,\,\,\,\,
\label{smallL3}\eea
For $h\gg\mpl/\xi$ (the inflationary regime) we can again switch back to the canonically normalized field $\heafi$, as we did earlier in eq.~(\ref{smallL2}), but now with the modified inflationary potential $V(\heafi)={\lambda\mpl^2\over 6\,\xi^2}\heafi^2 G(\heafi)$.

Otherwise, in case (ii), the inflationary regime is modified from (\ref{sigmaaction}) to the action
\beq
S=\int d^4x\sqrt{-g}\left[{1\over2}\mpl^2\mathcal{R}+{1\over 2}(\partial\heafi)^2
-{\newfi\mpl^2\over6\,\xi^2}\heafi^2 F\!\left(\heafi\right) \right]
\label{sigmaactionF}\eeq
In either case, since we have freedom to choose $G$, $F$, then not only is the scale of inflation altered, but in general we will not even have a quadratic chaotic inflation model. So essentially all of the inflationary predictions, such as the spectral index $n_s$, will be altered.

\section{Supergravity}\label{Supergravity}
There are various pieces of evidence to indicate that there exists new physics beyond the standard model, this comes from dark matter, baryogenesis, unification, hierarchy problem, non-renormalizability of gravity, etc. Arguably the most promising approach to address some or all of these issues is supersymmetry.
When gravity is included, global supersymmetry must be promoted to a local symmetry, known as supergravity. Hence, it is of interest to know if the framework for inflationary model building discussed in this paper can be incorporated into the framework of supergravity.

In order to do so, we need two extra complex scalar fields $S_1$ and $S_2$. We also need to promote $h$ to a complex field $H$ with Re$(H)=h/\sqrt{2}$ and $\heafi$ to a complex field $\heafiC$ with Re$(\heafiC)=\heafi/\sqrt{2}$.
We take the following superpotential and K\"ahler potential:
\bea
&&W=\sqrt{\lambda}\,S_1H^2+\sqrt{\newfi}\,S_2\!\left({\mpl\over\sqrt{3}\,\xi\,}\heafiC-H^2\right)\label{SuperpotentialUV}\\
&&K=-3\mpl^2\ln\!\left[1+\sum_i{(\psi^i-\bar{\psi}^i)^2\over6\mpl^2}\right]\label{KahlerUV}
\eea
where $\psi^i\equiv\{H,\Phi,S_1,S_2\}$.
The scalar sector of the Einstein frame action is given by the $\mathcal{N}=1$ supergravity action
\bea
S=\amp\amp\!\!\!\!\!\!\int d^4x\sqrt{-g}\Bigg{[}{1\over2}\mpl^2\mathcal{R}+\sum_{ij}K_{i\bar{j}}\partial\psi^i\partial\bar{\psi}^j\nonumber\\
\amp-\amp e^{K/\mpl^2}\Bigg{(}\sum_{ij}D_iWK^{i\bar{j}}\overline{D_jW}
-3W\overline{W}/\mpl^2\Bigg{)}\Bigg{]}\,\,\,\,\,\,\,\,\,\,\,
\label{SUGRA}\eea
Then by setting $\mbox{Im}(H)=\mbox{Im}(\heafiC)=S_1=S_2=0$ 
we find the action of eq.~(\ref{action}). Note that one should also include appropriate terms for the solution to be stable, such as a cubic term $S_1^3$, but we will not track such details here.

If $\lambda\ll \newfi$ then we can just use the low energy effective field theory. 
In this case, we only need a total of two fields: $H$ and $S\,(=S_1)$.
We take the following superpotential and K\"ahler potential:
\bea
W\amp=\amp\sqrt{\lambda}\,S H^2\label{Superpotential}\\
K\amp=\amp-3\mpl^2\ln\Bigg{[}1+{(H-\bar{H})^2\over6\mpl^2}+{(S-\bar{S})^2\over6\mpl^2}\nonumber\\
\amp\amp\;\;\;\;\;\;\;\;\;\;\;\;\;\;\;\;\;\;\;\;\;\;\;\;\;\;\;\;\;\;\;\;\; +{\xi^2(H^2-\bar{H}^2)^2\over2\mpl^4}\Bigg{]}\,\,\,\label{KahlerEFT}
\eea
Then by setting $\mbox{Im}(H)=S=0$  and using (\ref{SUGRA}) we find the action of (\ref{smallL1}), which furnishes a supergravity version of the simple model.

Lets compare this to the non-minimally coupled models studied elsewhere, 
such as Refs.~\cite{Einhorn:2009bh,Ferrara:2010yw,Ferrara:2010in}. 
One such version, which carries all the basic qualitative features of interest here, is to take the superpotential to be the same (eq.~(\ref{Superpotential})), while taking the K\"ahler potential to be 
\bea
K\amp=\amp-3\mpl^2\ln\Bigg{[}1+{(H-\bar{H})^2\over6\mpl^2}+{(S-\bar{S})^2\over6\mpl^2}\nonumber\\
\amp\amp\;\;\;\;\;\;\;\;\;\;\;\;\;\;\;\;\;\;\;\;\;\;\;\;\;\;\;\;\;\;\;\;\;\;\;\; +{\xi(H^2+\bar{H}^2)\over\mpl^2}\Bigg{]}
\eea
This K\"ahler potential may appear more fundamental than the minimally coupled model in (\ref{KahlerEFT}) since the argument of the logarithm is quadratic in $H$, rather than quartic. But this does not seem important from the effective field theory point of view, because in both cases a new mass scale has been introduced $\Lambda\sim\mpl/\xi$ (for large $\xi$), and the UV complete version in eq.~(\ref{KahlerUV}) is quadratic anyhow. So both models require new physics to enter before the scale $\Lambda\sim\mpl/\xi$ and so both appear equally arbitrary. It is conceivable that some top-down approach would prefer the argument of the logarithm to be quadratic, as opposed to other functions, though this does not appear to be the case in typical string models.

\section{Conclusions}\label{Conclusions}
In this paper we have studied a simple version of Higgs-inflation by minimally coupling all fields to gravity and examining a specific dimension 4 Lagrangian (eq.~(\ref{action})). We also exhibited a generalization to an entire class (eq.~(\ref{actionG})). 
Although we examined only minimally coupled models, our qualitative conclusions apply equally well to the non-minimally coupled models also. All such models require new physics to enter before the UV scale $\Lambda\sim\mpl/\xi$, and so, in a fashion
similar to Ref.~\cite{Giudice:2010ka}, we have introduced a second field to track the UV behavior. We found that the inflationary physics is controlled by the relative size of the self coupling $\lambda$ to the size of the coupling $\newfi$ in the new sector.

In case (i) (i.e., requiring $\lambda\ll\newfi\lesssim1$) the low energy effective field theory for $h$ (eq.~(\ref{smallL1}) or ({\ref{smallL3})) can be used to describe inflation. It is a simple minimally coupled model of inflation which differs from the non-minimally coupled models of \cite{Salopek,Bezrukov:2007ep,Giudice:2010ka}. 
It is a quadratic inflationary model due to the interplay between the $h^4$ and the $h^2(\partial h)^2$ terms in the effective field theory.
This shows that the Higgs-inflation models are rather arbitrary, as these are just two of many possibilities; a class of which comes from the functional freedom in $G(h)$. So the high energy physics of inflation is a matter of choice.
In such cases, we require $\lambda$ to be quite small in order for the low energy effective field theory governing the Higgs to be adequate.
In case (ii) (i.e., $g\ll\lambda$, which allows $\lambda$ to be moderately sized) the low energy effective field theory breaks down during inflation, in accordance with Ref.~\cite{Hertzberg:2010dc}. Instead it must be replaced by new degrees of freedom and the inflationary phase is entirely determined by the details of the new physics (eq.~(\ref{sigmaaction}) or ({\ref{sigmaactionF})), such as $\newfi$ and $F(\heafi)$, having little to do with the Higgs. In general this leads to altered cosmological predictions.
This arbitrariness in the choice of the high energy potential suggests that inflation is  disconnected from  
low energy physics. 

We embedded the model into a supergravity framework (eqs.~(\ref{SuperpotentialUV},\,\ref{KahlerUV})), although the basic field theory arbitrariness might not be fixed even in this setting. A full analysis of this deserves further investigation.

\section*{Acknowledgments}
We would like to thank Renata Kallosh and Andrei Linde for interesting discussion.
We acknowledge support by NSF grant PHY-0756174 and support from a Kavli Fellowship.

\end{document}